
\input harvmac.tex

\def\inbar{\,\vrule height1.5ex width.4pt depth0pt}

\def\Tr{\rm Tr}
\def\p{\partial}
\def\pb{\bar{\p}}
\font\cmss=cmss10 \font\cmsss=cmss10 at 7pt

\def\IR{\relax{\rm I\kern-.18em R}}

\def\Z{\relax\ifmmode\mathchoice
{\hbox{\cmss Z\kern-.4em Z}}{\hbox{\cmss Z\kern-.4em Z}}
{\lower.9pt\hbox{\cmsss Z\kern-.4em Z}}
{\lower1.2pt\hbox{\cmsss Z\kern-.4em Z}}\else{\cmss Z\kern-.4em
Z}\fi}
\def\IB{\relax{\rm I\kern-.18em B}}
\def\C{{\relax\hbox{$\inbar\kern-.3em{\rm C}$}}}
\def\ID{\relax{\rm I\kern-.18em D}}
\def\IE{\relax{\rm I\kern-.18em E}}
\def\IF{\relax{\rm I\kern-.18em F}}
\def\P{\relax{\rm I\kern-.18em P}}
\def\IG{\relax\hbox{$\inbar\kern-.3em{\rm G}$}}
\def\IGa{\relax\hbox{${\rm I}\kern-.18em\Gamma$}}
\def\IH{\relax{\rm I\kern-.18em H}}
\def\II{\relax{\rm I\kern-.18em I}}
\def\IK{\relax{\rm I\kern-.18em K}}
\def\neib{neighborhood}

\tolerance 20000

\hfill\vtop{\baselineskip12pt\hbox{hep-th/9503157}
\hbox{PUPT-1534}
\hbox{ITEP-N95/1}}

\bigskip
\bigskip
\centerline{\bf HOLOMORPHIC BUNDLES AND MANY-BODY SYSTEMS}

\vskip 1cm
\centerline{\rm Nikita Nekrasov}
\vskip 1cm
\centerline{\it Department of Physics,
Princeton University, Princeton NJ 08544}
\vskip 0.1cm
\centerline{and}
\vskip 0.1cm
\centerline{\it Institute of Theoretical and Experimental Physics,
117259, Moscow, Russia}
\vskip 1cm
\centerline{e-mail: nikita@puhep1.princeton.edu}
\bigskip
\bigskip
    We show that spin generalization of elliptic
Calogero-Moser system, elliptic extension of Gaudin model
and their cousins can be treated as a degenerations
of Hitchin systems. Applications to the constructions
of integrals of motion,
angle-action variables and quantum
systems are discussed.
\medskip
The constructions are motivated by the Conformal Field Theory, and
their quantum counterpart
can be treated as a degeneration of the critical level
Knizhnik-Zamolodchikov-Bernard equations.

\vfill\eject

\newsec{\bf Introduction}
    Integrable many-body systems attract attention for the following
reasons: they are important in condensed matter physics and they appear
quite often in two dimensional gauge theories
as well as in conformal field theory. Recently
they have been recognized in four dimensional gauge theories.

Among these systems the following
ones will be of special interest for us:
{\item 1.} {\it Spin generalization of Elliptic Calogero - Moser model} -
it describes the system of particles in one (complex) dimension,
interacting through the pair-wise potential.
The explicit form of the Hamiltonian is:
$$
H = \sum_{i=1}^{N} {p_{i}^{2}\over{2}}  +
\sum_{i \neq j} {\Tr} (S_{i} S_{j}) {\wp}(z_{i}-z_{j})
$$
where $z_{i}$ are the positions of the particles,
$p_{i}$ - corresponding
momenta and $S_{i}$ are the "spins" - $l \times l$ matrices, acting in some
auxilliary space.
The conditions on $S_{i}$ will be specified later.
The only point to be mentioned is that the Poisson brackets between
$p, z, S$ are the following:
$$
\{ p_{i}, z_{j} \} = \delta_{ij}
$$
$$
\{ (S_{i})_{ab}  , (S_{j})_{cd} \} =
{\delta_{ij}} ( {\delta}_{ad} (S_{i})_{bc}- {\delta}_{bc} (S_{i})_{ad})
$$
{\item 2.} {\it Gaudin model and its elliptic counterpart}.
We describe first
the rational case.
Consider a collection of $L$ points on ${\P}^{1}$ in
generic position: $w_{1}, \dots, w_{L}$,
assign to each $w_{I}$ a spin $S_{I}$ (an $N \times N$ matrix) and
define the Hamiltonians [G]:
$$
H_{I} = \sum_{J \neq I} {{{\Tr} ( S_{I} S_{J} ) }\over{(w_{I} - w_{J})}}
$$

The main goal of this
note is to include these two (seemingly) unrelated models
in the universal family of integrable models,
naturally related to the moduli spaces of
holomorphic bundles over the curves.
  It will turn out, that the appropriate
objects to study are Hitchin systems.
As a by-product we shall invent elliptic Gaudin model, which
includes both cases as a special limits.
We shall also obtain a
prescription for construction
of integarls of motion and action-angle variables.
The paper is organized as follows.
In the section {\bf 2.} we remind the construction of Hitchin systems.
The section {\bf 3.} is devoted to the explanation of
the mapping between the Hitchin systems and the models, just described.
The section {\bf 4.} deals with action-angle variables and
integrals of motion.
We conclude with the remarks on the quantization
of our constructions.

\newsec{Acknowledgements}

I would like to thank Sasha Gorsky for collaboration in [NG1],[NG2] and
[NG3], A. Losev and D. Ivanov for useful discussions,
V. Fock,
I. Frenkel, G. Moore, A. Polyakov,
M. Olshanetsky, A. Rosly , V. Rubtsov and  S. Shatashvili for kind advises.

\newsec{ Construction of the  Systems}

We must confess that  all the models we are discussing are motivated
by the studies of Knizhnik-Zamolodchikov-Bernard equations [KZ],[Be],
[Lo],[FV],[I]. Our paper is a development of [NG1], [NG2] .
One of the outcomes of our work might be an insight in the ${\cal W}$-
generalizations of them.

\subsec{ Hitchin systems}
Hitchin has introduced in  [H] a family of integrable systems. The phase
space of these systems can be identified with the cotangent bundle
$T^{*}{\cal N}$ to the moduli space $\cal N$ of
stable holomorphic vector bundles of rank $N$
(for $GL_{N}({\C})$ case) over the compact
smooth Riemann surface $\Sigma$ of genus $g > 1$. His construction can
be briefly described as follows.
Fix the topological class of the bundles (i.e. let us
consider the bundles $\cal E$ with $c_{1}({\cal E}) = k$, with $k$ -
fixed).
Consider the space ${\cal A}^{s}$ of stable complex structures
in a given smooth vector bundle $V$,
whose fiber is isomorphic to ${\C}^{N}$.
The notion of stable bundle comes from geometric invariants
theory and implies in this context, that for  any proper
subbundle $U$:
$$
{{deg (U)}\over{rk(U)}} < {{deg(V)}\over{rk(V)}}
$$
The quotient of ${\cal A}^{s}/{\cal G}$ of the space of
all stable complex structures by the gauge group is
the moduli space $\cal N$.
Its dimension is given by the Riemann-Roch theorem
$$
dim({\cal N}) = N^{2}(g -1) + 1
$$
Now consider a cotangent bundle to ${\cal A}^{s}$. It is the space
of pairs:
$$
\phi, d_{A}^{\prime \prime}
$$
where
$\phi$ is a $Mat_{N}({\C})$ valued $(1,0)$ - differential on $\Sigma$,
$d_{A}^{\prime \prime}$ is an
operator, defining
the complex structure on $V$:
$$
d_{A}^{\prime \prime} : \Omega^{0} ({\Sigma}, V)
\to \Omega^{0,1} ({\Sigma}, V)
$$
The field $\phi$ is called a Higgs field
and the pair $d_{A}^{\prime \prime}, \phi$ defines what
is called a Higgs bundle. In the framework of conformal field theory the Higgs
field is usually reffered to as the holomorphic current, while holomorphic
bundle
defines a background gauge field.

The cotangent bundle $T^{*}{\cal A}^{s}$
can be endowed with a holomorphic symplectic form:
$$
\omega = \int_{\Sigma}
{\Tr} \delta \phi \wedge \delta d_{A}^{\prime \prime}
$$
where
$\delta d_{A}^{\prime \prime}$ can be identified
with a $(0,1)$ - form with values in $N$ by $N$ matrices.
Gauge group $\cal G$ act on $T^{*}{\cal A}^{s}$ by the
transformations:
$$
\phi \to g^{-1} \phi g
$$
$$
d_{A}^{\prime \prime} \to g^{-1} d_{A}^{\prime \prime} g
$$
and preserves the form $\omega$.
Therefore, a moment map is defined:
$$
\mu = [ d_{A}^{\prime \prime}, {\phi} ]
$$
Taking the zero
level of the moment map and
factorizing it along the orbits of
$\cal G$ we get
the symplectic quotient,
which can be identified with
$T^{*}{\cal N}$.
Now the Hitchin Hamiltonians are
constructed with the help of
holomorphic $(1-j,1)$-
differentials
$\nu_{j,i_{j}}$
where $i_{j}$
labels a basis in the
linear space
$$
H^{1} ( {\Sigma} , {\cal K} \otimes {\cal T}^{j})  = {\C}^{(2j-1)(g-1)}
$$
for
$j > 1$ and ${\C}^{g}$ for $j = 1$.
Take a gauge invariant $(j,0)$-differential
${\Tr} {\phi}^{j}$
and integrate it over
$\Sigma$ with the weight
$\nu_{j,i_{j}}$:
$$
H_{j, i_{j}} = \int_{\Sigma} \nu_{j,i_{j}} {\Tr} {\phi}^{j}
$$
Obviously, on
$T^{*}{\cal A}^{s}$
these functions
Poisson-commute.
Since they are gauge invariant,
they will Poisson-commute
after reduction.
Also it is obvious,
that they are functionally
independent and their
total number is equal to
$$
g + \sum_{j=2}^{N} (  2j -1 ) (g -1) = N^{2} (g -1) + 1 = dim({\cal N})
$$
Therefore,
we have an integrable system.
\subsec{\bf Holomorphic bundles
over degenerate curves}
Now let us consider a degeneration of the curve.
Recall, that the normalization
of the stable curve $\Sigma$ is a collection of a
smooth curves $\Sigma_{\alpha}$
with possible marked points,
such that any component of genus
zero has at least three marked
points and every component of genus one
has at least one such a point.
For each component $\Sigma_{\alpha}$
we have a subset $X_{\alpha} = \{ x_{\alpha}^{1},
\dots,x^{L_{\alpha}}_{\alpha} \}$ of points. Let us denote
the pair $({\Sigma}_{\alpha}, X_{\alpha})$ as $C_{\alpha}$.
The disjoint
union of $C_{\alpha}$'s is mapped
onto $\Sigma$ by the normalization map $\pi$.
Let us denote by $X_{\alpha \beta}$ the
set of double points
${\pi}(X_{\alpha}) \cap {\pi}(X_{\beta})$ for $\alpha \neq \beta$ and
as $X_{\alpha \alpha}$ the set of double points in ${\pi}(X_{\alpha})$ (these
appear due to pinching the handles). The union
of all $X_{\alpha \beta}$ we shall denote by $X \subset {\Sigma}$.
We define $x_{\alpha \beta}^{ij} \in X_{\alpha \beta}$
as ${\pi}(x_{\alpha}^{i})\cap {\pi} (x_{\beta}^{j})$.
Notice, that it may be empty.
{\it Stable bundle $\cal E$ over $\Sigma$} is a collection of
holomorphic bundles
${\cal E}_{\alpha}$ over $\Sigma_{\alpha}$
of rank $N$
(there might be some
generalizations with different ranks
of the bundle over different components -
these are
unnatural as
a degeneration of the bundle over smooth curve)
with the
identifications
$g_{\alpha \beta}^{ij}$
of the fibers
$$
g_{\alpha \beta}^{ij} :
{\cal E}_{\alpha} |_{x_{\alpha}^{i}}\to
 {\cal E}_{\beta} |_{x_{\beta}^{j}}
$$
with the obvious condition:
$g_{\alpha \beta}^{ij} g_{\beta \alpha}^{ji} = 1$.

Gauge group acts on the complex structure of the bundle
${\cal E}_{\alpha}$
for each $\alpha$
as in the smooth curve case.
The new ingredient is the action on $g_{\alpha \beta}^{ij}$.
Fix a gauge
transformations
$g_{\alpha}$
for each component of
$\Sigma$.
Then
$g_{\alpha \beta}^{ij}$
are acted on by
$g_{\alpha}$ as follows:
$$
g_{\alpha \beta}^{ij}
\to g_{\beta}(x_{\beta}^{j})^{-1} g_{\alpha \beta}^{ij} g_{\alpha}(
x_{\alpha}^{i})
$$
Now we have to introduce a notion of stable bundle.
The condition  of stabilty is:

{\it For each collection of proper  subbundles
${\cal F}_{\alpha} \subset {\cal E}_{\alpha}$, such that
$$
g_{\alpha \beta}^{ij}
( {\cal F}_{\alpha} |_{x_{\alpha}^{i}}) =
{\cal F}_{\beta} |_{x_{\beta}^{j}}
$$
and
$$
rk ({\cal F}_{\alpha}) = N^{\prime} < N
$$
for each
$\alpha$
the following inequality holds:
$$
deg( {\cal F}_{\alpha} ) < {N^{\prime}\over{N}} deg({\cal E}_{\alpha})
$$
for any
$\alpha$.}

Let
${\cal A}$ will denote space of collections of
$d_{A, \alpha}^{\prime \prime}$ operators in each ${\cal E}_{\alpha}$
together with $g_{\alpha \beta}^{ij}$ for each $\alpha$ and $\beta$.
Let ${\cal A}^{s}$ will denote the subspace of $\cal A$,
consisting of
{\it stable} objects.
The cotangent bundle $T^{*}{\cal A}^{s}$ can be identified
with the space of collections of pairs
$$
({\cal E}_{\alpha}, {\phi}_{\alpha}) , {\phi}_{\alpha} \in
\Omega^{1,0}({\Sigma}_{\alpha})
\otimes End( {\cal E}_{\alpha})
$$
and
$$
(g_{\alpha \beta}^{ij}, p_{\alpha, \beta}^{ij}),
p_{\alpha \beta}^{ij} \in
T^{*}_{g_{\alpha \beta}^{ij}}
Hom( {\cal E}_{\alpha}|_{x_{\alpha}^{i}},
{\cal E}_{\beta}|_{x_{\beta}^{j}} )
$$
We normalize
$p_{\alpha \beta}^{ij}$:
$p_{\alpha \beta}^{ij} =
- Ad^{*}(g_{\alpha \beta}^{ij}) p_{\beta \alpha}^{ji}$.
The Higgs
fields
$\phi_{\alpha}$
are allowed to have singularities at the marked
points.
As we will see,
they could have poles there.
Now we shall proceed as in the previous section.
Consider the gauge
group action on
$T^{*}{\cal A}^{s}$.
Since the gauge group
$\cal G$ is
essentially the product of gauge groups
${\cal G}_{\alpha}$,
the moment map
is a collection of
the moment maps
for each component
$\Sigma_{\alpha}$:
$$
\mu_{\alpha} = [ d_{A, \alpha}^{\prime \prime} ,
{\phi}_{\alpha} ] +
\sum_{\beta, i, j} p_{\alpha \beta}^{ij}
\delta^{2}(x_{\alpha}^{i})
$$
where the
sum over $i$ runs from $1$ up to
$L_{\alpha}$
while
$\beta$ and $j$ are determined from the condition, that
${\pi}(x_{\beta}^{j})
= {\pi}(x_{\alpha}^{i})$.
Let us now repeat the procedure of reduction.
At the first step we
should restrict ourself onto the zero level of the moment map.
It means, that $\phi_{\alpha}$ becomes a meromorphic section
of the bundle $End({\cal E}_{\alpha})
\otimes \Omega^{1,0}({\Sigma}_{\alpha})$
with the first order poles at the double points. The residue
of $\phi_{\alpha}$  at the point $x_{\alpha}^{i}$ equals to
$p_{\alpha \beta}^{ij}$
for appropriate ${\beta}, j$.
This condition is compartible with the definition
of the canonical bundle over the stable curve.
On the next step we take a quotient with respect to the
gauge group action and get the reduced space
$T^{*}{\cal N}$.
The space $\cal N$ is the quotient of ${\cal A}^{s}$ by $\cal G$.
The symplectic form on $T^{*}{\cal N}$ can be written
as:
$$
\omega = \sum_{\alpha} \omega_{\alpha} +
\sum_{({\alpha},i), ( {\beta},j)} {\Tr} \delta (g_{\beta \alpha}^{ji}
p_{\alpha \beta}^{ij} )
\wedge
\delta g_{\alpha \beta}^{ij}
$$
Let us calculate the dimension of
$T^{*}{\cal N}$. We shall
calculate the (complex) dimension of ${\cal N}$ by
means of the following trick.
The moduli space $\cal N$ can be projected onto the direct product
of moduli spaces ${\cal N}_{\alpha}$ of the stable bundles
over ${\Sigma}_{\alpha}$'s.
Actually, the map is to the product of the moduli of
holomorphic bundles,
but the open dense subset,
consisting of the stable bundles is covered.
The projection simply takes the collection
of ${\cal E}_{\alpha}$'s
to the product of equivalence
classes in $ {\cal N}_{\alpha}$'s.
The fiber of this map can be identified
 with the quotient $G/H$, where $G$ is the
set of collections of $g_{\alpha \beta}^{ij}$,
while $H$ is the group of automorphismes of $\times_{\alpha}
{\cal E}_{\alpha}$. This
group is a product over all
components $\Sigma_{\alpha}$ of genus
$g({\Sigma_{\alpha}}) < 2$ zero
of
the groups $H_{\alpha}$.
For genus zero component $H_{\alpha}$
is $GL_{N}({\C})$, while genus one component
provides a maximal torus - $({\C}^{*})^{N}$.
Therefore, at generic point, we conclude,
that the dimension of
$\cal N$ is
$$
dim({\cal N}) = \sum_{\alpha} dim({\cal N}_{\alpha}) + dim(G/H)  =
N^{2} E({\Sigma}) +
\sum_{\alpha} N^{2}( g({\Sigma}_{\alpha}) - 1) = N^{2}(g-1) + 1
$$
where we
have used Riemann-Roch theorem in the form
$$
dim({\cal N}_{\alpha})
- dim(H_{\alpha})
= N^{2} (g({\Sigma}_{\alpha}) - 1),
$$
$E({\Sigma})$ is the total number
of double points.
\subsec{\bf Hamiltonian systems on $T^{*}{\cal N}$}
Now we shall define the Hamiltonians.
For each $\alpha$ we take
${\nu}_{\alpha, l,k}$ - the $k$'th holomorphic $(1-l,1)$ differential
on ${\Sigma}_{\alpha} - X_{\alpha}$ and construct  a holomorphic function
on $T^{*}{\cal A}^{s}$:
$$
H_{{\alpha},l,k} = \int_{\Sigma_{\alpha}} \nu_{{\alpha},l,k}
{\Tr} ({\phi}_{\alpha}^{l})
$$
Obviously, all $H_{{\alpha}, l,k}$ descend to $T^{*}{\cal N}$ and
Poisson-commute
there.

One can notice, that the integrable systems we get can be restricted
onto the invariant submanifolds. Namely, the
conjugacy classes of $p_{\alpha \beta}^{ij}$ are invariant under the flows,
which we  have constructed. Indeed, the casimirs
${\Tr} (p_{\alpha \beta}^{ij})^{l}$ are the coefficients at
the most singular terms
of ${\Tr} {\phi}^{l}$, therefore, they are among the Hamiltonians we have
constructed.

\newsec{\bf Gaudin model, Spin Elliptic Calogero-Moser System and so on ...}
\subsec{\bf Genus zero models}
Consider a component of genus zero.
 Let us describe explicitely the part
of $T^{*}{\cal N}$ related to this component as well as the Hamiltonians.
We shall omit the label
$\alpha$ in the subsequent formulas to save the print.
Thus, we have $C = ({\P}^{1}, x_{1}, \dots, x_{L})$.
\centerline{\it Gaudin model}
Assume first, that for $i \neq j$:
${\pi}(x_{i}) \neq {\pi}(x_{j})$.
Then $p_{\alpha \beta}^{ij}$ can
be denoted simply as $p_{i}$ without any confusion.
There are no moduli of holomorphic bundles over the sphere,
therefore,
we can
assume
that $d^{\prime \prime}_{A}$
is just the $\pb$ operator in the line bundle $\cal L$ of
the degree $deg({\cal E})$.
The moment map condition:
$$
0
= {\pb} \phi + \sum_{i} p_{i} \delta^{2}(x_{i})
$$
is
easily solved:
$$
\phi (x) = - {1\over{2{\pi}{\sqrt{-1}}}}
\sum_{i} {p_{i}\over{(x - x_{i})}}
$$
Notice,
however, that every holomorphic bundle over
${\P}^{1}$ has
an automorphism group $GL_{N}({\C})$,
which acts nontrivially
on $\phi$ as well as on $p_{i}$.
In fact, the reduction
with respect to this subgroup is forced by our equation:
the sum of
residues of $\phi$ must vanish, giving rise to the constraint:
$$
\sum_{i} p_{i} = 0
$$
which is nothing but the moment map
for the $GL_{N}({\C})$ action.

Our Hamiltonians in this case boil down to

$$
H_{l,a,b}
= Res_{x_{a}} (x - x_{a})^{b-1} {\Tr} {\phi}^{l}(x),
$$
where $1 \leq b \leq  l, 1 \leq a \leq L$.
These
Hamiltonians (essentially $H_{2,a,2}$)
are called {\it Gaudin Hamiltonians}.

$$
H_{1,a,1} = {\Tr} (p_{a}),
H_{2,a,1} = {\Tr} (p_{a}^{2}) ,
$$
$$
H_{2,a,2} =
\sum_{b \neq a}
{{{\Tr} (p_{b} p_{a})}\over{(x_{b} - x_{a})}}
$$
etc.

\centerline{\it Spin Calogero-Moser and Rational Ruijsenaars Systems}

Now consider genus zero component
$\Sigma_{\alpha}$ with double
points.
Let us decompose the set of marked points
$X_{\alpha}$ as
$$
X_{\alpha} = S \cup T \cup {\sigma} (T)
$$
where
$t \in T$ and
${\sigma}(t) \in {\sigma}(T)$
are mapped to $x_{\alpha \alpha}^{t {\sigma}(t)}$,
while the restriction of
$\pi$ on$S$ is surjective.
We denote
$p_{\alpha \alpha}^{t {\sigma}(t)} = p_{t}$,
$g_{\alpha \alpha}^{t {\sigma}(t)} = g_{t}$.
We have: $p_{\alpha \alpha}^{{\sigma}(t) t} = - Ad^{*}(g_{t}) p_{t}$.
Solving the moment map condition
as before, we get:
$$
\phi_{\alpha}(x) = \sum_{s} {p_{s}\over{(x - x_{s})}} +
\sum_{t} { p_{t}\over{(x - x_{t})}}
- {{Ad^{*}(g_{t}) p_{t}}\over{(x - x_{{\sigma}(t)})}}
$$
Sum of the residues vanishes:
$$
\sum_{s} p_{s} + \sum_{t} (p_{t} - Ad^{*}(g_{t}) p_{t}) = 0
$$
The group
$GL_{N}({\C})$ of
automorphismes acts on the data
$p_{s}, p_{t}, g_{t}$
as follows:
$$
g_{t} \to g^{-1} g_{t} g, p_{t} \to g^{-1} p_{t} g, p_{s}
\to g^{-1} p_{s} g
$$

Now let us specialize to the case, when $\# T = 1$. The moment map
condition will give
$$
p_{t} - Ad^{*}(g_{t})p_{t} = - \sum_{s} p_{s}
$$
We have two options:

we can parametrize the quotient by the action of the group
$GL_{N}({\C})$ either by
fixing the conjugacy class of $g_{t}$, or the one of $p_{t}$.

Consider the first option. Generically one can diagonalize
$g_{t}$, and there will be left a group of diagonal matrices, which
will act nontrivially on $p_{t}$ and $p_{s}$'s.
Let us denote the eigenvalues of $g_{t}$ by
$$
g_{t} \sim diag( e^{z_{1}}, \dots , e^{z_{N}} )
$$
Then in the basis, where $g_{t}$ is diagonal, $p_{t}$ has
a form:
$$
(p_{t}){ij} = p_{i} \delta_{ij} + {{ \sum_{s} (p_{s})_{ij} }\over{1-
e^{z_{i}-z_{j}}}}
$$
with the further condition
$$
\sum_{s} (p_{s})_{ii} = 0
$$
for any $i$. This condition has an elliptic nature, as we will see, and
has a very natural origin: the double point on the sphere
comes from the pinching the handle.

Explicitely calculating $H_{2,t,2}$ we
get:
$$
\sum_{i} {p_{i}^{2}\over{2}} + \sum_{i,j}
{{S_{ij}}\over{sinh^{2}({{z_{i}-z_{j}}\over{2}} )}}
$$
with
$$
S_{ij} = \sum_{s,s^{\prime}} (p_{s})_{ij} (p_{s^{\prime}})_{ji}
$$

This Hamiltonian describes the particles with spin
interaction. The way spins come out will be clear later. Right now we
simply claim that $S_{ij}$ is the relevant spin interaction.

Now let us investigate another option - namely, we diagonalize
$p_{t}$. For simplicity we shall treat the case $\# S =1$.
We have:
$$
p_{t} = diag( {\theta}_{1}, \dots, {\theta}_{N} )
$$
and
$$
(g_{t})_{ij} ({\theta}_{i} - {\theta}_{j}) = ({\tilde p}_{s})_{ij}
$$
where ${\tilde p}_{s} = g_{t} p_{s}$.
Now we make a further restriction. Suppose, that for some $\nu \in {\C}^{*}$
$p_{s} - {\bf Id} {{\nu}}$ has rank one.

Then,
$$
p_{s} = \nu {\bf Id} + u \otimes v
$$
where $v \in ({\C}^{N})^{*}, u \in {\C}^{N}$.

Therefore, we can solve for $g_{t}$:
$$
(g_{t})_{ij} = {{{\tilde u}_{i} v_{j}}\over{
{\theta}_{i} -
{\theta}_{j} - {\nu}}}
$$
and $\tilde u_{i} = (g_{t}u)_{i}$.
Thus, we obtain a linear equation:
$$
\sum_{j} {{u_{j}v_{j}}\over{{\theta}_{i} - {\theta}_{j} - {\nu}}} = 1
$$
for any $i$. The solution is
$$
u_{i}v_{i} = {{{\P}({\theta}_{i} + {\nu})}\over{{\nu}
{\P}^{\prime}({\theta}_{i})}}
$$
where ${\P} ( {\theta} ) = \prod ( {\theta} - {\theta}_{i} )$.

Finally we introduce the coordinates $z_{i}$, defined by
$$
e^{z_{i}} u_{i} = {\tilde u}_{i}
$$

The Hamiltonians we can consider in this approach are the characters
of $g_{t}$.
We have:
$$
{\Tr} g_{t}^{k} = \sum_{I \subset \{ 1, \dots , N \}, \# I = k}
e^{z_{I}} \prod_{i \in I, j \not\subset I}
{{{\theta}_{i} - {\theta}_{j} + {\nu}}\over{{\theta}_{i} - {\theta}_{j}}}
$$

The system we get is called {\it Rational Macdonald} system. If, instead
of taking $G = GL_{N}({\C})$ we would consider $SU(N)$, we will
get what is caled a rational Ruijsenaars model [R], [RS], [NG2].

\subsec{\bf Elliptic models}

The next interesting example is the
genus one component.
Again we omit label $\alpha$ and
$p_{\alpha \beta}^{ij}$ gets replaced
by $p_{i}$.
Generic holomorphic bundles over the torus are decomposable
into the direct sum of the line bundles:
$$
{\cal E} = \oplus_{i=1}^{N} {\cal L}_{i}
$$
Therefore,
the moduli space
${\cal N}_{\alpha}$
can be identified
with the $N$'th power of the
Jacobian of the curve, divided
by the permutation group action.
 Let us introduce the coordinates
$z_{1}, \dots, z_{N}$ on ${\cal N}_{\alpha}$.
They are defined up to the elliptic affine Weyl group action.
Let $\tau$ be the modular parameter of
the elliptic curve. Then
there are shifts
$$
z_{a} \to z_{a} + 2{\pi}{\sqrt{-1}}{{m_{a} {\tau} +
n_{a}}\over{{\tau} - {\bar \tau}}}
$$
with
$m_{a},n_{a}\in {\Z}$,
induced by the gauge transformations
$$
\exp( diag( 2{\pi}{\sqrt{-1}}
{{n_{a}( x - {\bar x} )
+ m_{a} ( x{\bar \tau} - {\bar x} {\tau} )}\over{{\tau} - {\bar \tau}}} ))
$$
as
well as
permutations of $z_{i}$'s.
Up to these equivalences $z_{i}$'s are the honest  coodinates.

\centerline{\it No double points}

First,
we dispose of the case,
when $\pi$ doesn't map two points to one.
Now the moment map condition is
$$
\pb \phi_{ij} + ( z_{i} - z_{j} ) \phi_{ij} +
\sum_{a} (p_{a})_{ij} \delta^{2}(x_{a}) = 0
$$
The solution of this equation yields a Lax operator
$\phi$:
for
$a \neq b$
$$
\phi_{ij} =
{{\exp(z_{ij} {{x - {\bar x}}\over{{\tau} -
{\bar \tau}}})}\over{2{\pi}{\sqrt{-1}}}}\sum_{a} (p_{a})_{ij}{{{\sigma}
(z_{ij} + x - x_{a})}\over{{\sigma}(z_{ij}){\sigma}(x - x_{a})}}
$$
where we have denoted for brevity:
$z_{ij} = z_{i} - z_{j}$.
For the
diagonal
components
we get
$$
\phi_{ii} (x) =  w_{i} + \sum_{a} (p_{a})_{ii} \zeta ( x - x_{a} )
$$
In these formulas
$\sigma$ and $\zeta$ are Weierstrass elliptic functions for the curve with
periods
$1$ and $\tau$.
The condition of vanishing residues sum is
$$
\sum_{a} (p_{a})_{ii} = 0
$$
for any
$a = 1, \dots, N$.
Now we can compute our Hamiltonians.
We have:
$$
4{\pi}^{2} {\Tr} {\phi}^{2}  (x) =
$$
$$
=\sum_{i} (w_{i} + \sum_{a} (p_{a})_{ii} \zeta ( x - x_{a} ))^{2}
-
\sum_{i \neq j; a,b} (p_{a})_{ij} (p_{b})_{ji}
{{{\sigma}(z_{ij} + x - x_{a}) {\sigma} (z_{ji} + x - x_{b})}\over
{{\sigma}(z_{ij})^{2} {\sigma} (x - x_{a}) {\sigma} (x - x_{b})}}
$$

Expanding this expression as:
$$
4{\pi}^{2} {\Tr} {\phi}^{2}  (x) =
(\sum_{a} {\wp} (x - x_{a}) H_{2,2,a} + {\zeta}( x - x_{a} ) H_{2,1,a} )
+ H_{2,0}
$$
we obtain:

$$
H_{2,2,a} = {\Tr} p_{a}^{2}
$$
as it was expected,
$$
H_{2,1,a} = \sum_{i} w_{i} (p_{a})_{ii} +
\sum_{b \neq a; i} (p_{a})_{ii}(p_{b})_{ii} {\zeta} (x_{a}-x_{b}) +
$$
$$
+
\sum_{b \neq a; i \neq j} e^{z_{ij}(x_{a}-x_{b})}
(p_{b})_{ij} (p_{a})_{ji}
{{{\sigma}(z_{ij} + x_{a} - x_{b})}\over
{{\sigma}(z_{ij}) {\sigma}(x_{a} - x_{b})}}
$$
These Hamiltonians we shall name as {\it Elliptic Gaudin Hamiltonians}.

The next interesting one is:
$$
H_{2,0} = \sum_{i} w_{i}^{2} +
\sum_{i \neq j;a} (p_{a})_{ij} (p_{a})_{ji} {\wp} (z_{i} - z_{j})
+ \sum_{i; a \neq b} (p_{a})_{ii} (p_{b})_{ii} ( {\wp} (x_{a} - x_{b}) -
{\zeta}^{2}(x_{a}-x_{b}) ) +
$$
$$
\sum_{i \neq j;a \neq b} e^{z_{ij}(x_{a}-x_{b})}
(p_{a})_{ij} (p_{b})_{ji}
{{{\sigma}(z_{ij} + x_{a} - x_{b})}\over
{{\sigma}(z_{ij}) {\sigma}(x_{a} - x_{b})}}
( {\zeta} ( x_{a} - x_{b} + z_{ij} ) - {\zeta} ( z_{ij}) )
$$

Higher Hamiltonians provide us with the integrals of motion of this
model.

\centerline{\it Double points}

In the case, when there are the double points, the
formulas for the Lax operator and Hamiltonians are nearly the same, the only
difference is in the conditions on the $p_{a}$'s.

\subsec{Spins and coadjoint orbits}

In this section we shall map the notations $p_{a}$ for the Lie
algebra elements to the spin notations $S_{i}$, which were
introduced in the begining of the paper.

\centerline{\it No double points}
For simplicity, we first consider the case without double points
on the component $C_{\alpha}$.
First of all, we shall fix the conjugacy
classes of all $p_{a}$'s, that is we restrict ourselves onto
the invariant subvariety:
$$
Tr p_{a}^{n} = \rho_{a,n} = const
$$
Therefore, each $p_{a}$ will represent a point on
the coadjoint orbit $O_{a}$ of $SL_{N}({\C})$. Again for simplicity
we assume, that this coadjoint orbit is of the generic type.
One can represent it as follows.

Fix a point $x_{a}$ and denote $p_{a}$ simply as $p$.
Introduce a sequence of vector spaces
$$
{\cal E}^{r} \subset \dots \subset {\cal E}^{0}
$$
and consider the space of operators
$$
U_{i}: {\cal E}^{i} \to {\cal E}^{i+1}, V_{i}: {\cal E}^{i+1} \to
{\cal E}^{i}
$$
with the canonical symplectic form:
$$
\sum_{i} Tr \delta U^{i} \wedge \delta V^{i}
$$
This form is invariant under the action of the group
$$
H = \times_{i=1}^{r} GL({\cal E}^{i})
$$
by the changes of bases. Therefore, one can make a Hamiltonian
reduction at some central level of the moment map. Formally
it amounts to imposing the constraints:
$$
U_{r-1} V_{r-1} = \zeta_{r} {\bf Id}_{{\cal E}^{r}}
$$
$$
U_{i-1} V_{i-1} - V_{i} U_{i} = \zeta_{i} {\bf Id}_{{\cal E}^{i}}
$$
for $i=2, \dots, r-1$. Here the complex numbers
$\zeta_{i}$ are related to the eigenvalues $\lambda_{i}$ of the
matrix $p$ via:
$$
\lambda_{i} - \lambda_{i-1} = \zeta_{i}
$$
the multiplicity of $\lambda_{i}$ equals $dim{\cal E}^{i} -
dim{\cal E}^{i+1}$.
Finally, $p = V_{0}U_{0}$.

now for each point $x_{a}$ we have a pair of matrices
$U_{a}, V_{a}$, such that $p_{a} = V_{a}U_{a}$.

Then
$$
(S_{i})^{ab} =
(U_{a})_{i} \otimes (V_{b})^{j}
\in {\cal E}^{1}_{a} \otimes {\cal E}^{1,*}_{b}
$$

Therefore, in our formulas for Hamiltonians we could replace
$$
(p_{a})_{ij} (p_{a})_{ji} \to Tr_{{\cal E}^{1}_{a}} S_{i} S_{j}
$$

Of course, for the products $p_{a}p_{b}$ there is no such
interpretation. Thus, $S_{I}$ operators in the Gaudin model are
just the matrices $p_{a}$ .

\newsec{Action-Angle Variables}

We first recall the construction of Hitchin in the case of the compact curve of
genus $g > 2$.

Given a point in the moduli space of Higgs bundles one can
construct a {\it spectral curve}
$S \subset {\P} ( T^{*}{\Sigma} \oplus 1 )$:
$$
{\cal R} (x , \lambda) = det ({\phi} (x) - {\lambda})
$$
where $\lambda$ is a linear coordinate on the fiber of the
cotangent bundle $T^{*}{\Sigma}$. This curve is well-defined, since
the equation, which defines it, is gauge invariant.

The curve $S$ is $N$-sheeted ramified covering of $\Sigma$, its genus
can be computed by the adjunction formula or using Riemann-Gurwitz theorem.
$$
g(S) = N^{2} (g - 1) + 1
$$
which agrees with the dimension of the moduli space of stable bundles
over $\Sigma$. Denote by $p$ the projection $S \to {\Sigma}$.

Given a stable bundle $\cal E$ over $\Sigma$ we can pull it back onto
$S$. There is a line subbbundle ${\cal L} \subset p^{*}{\cal E}$, whose
fiber at generic point $(x,{\lambda})$ is an eigenspace
of $\phi (x)$ with the eigenvalue $\lambda$. Conversely,
given a line bundle $\cal L$ on $\Sigma$, on can take its
direct image, which (again at generic point) is defined as
$$
{\cal E}_{x}  = \oplus_{y \in p^{-1}(x)} {\cal L}_{y}
$$

Therefore, under the flow, generated by the Hitchin Hamiltonians,
the ${\cal L}$ changes and it can be shown, that these flows extend to the
linear commuting vector fields on the Jacobian $Jac(S)$ of $S$.

Thus, the linear coordinates on $Jac(S)$ are the coordinates of the
angle-type, whereas the intregrals of
${\lambda}$ over the corresponding cycles in $S$ give the action variables.

The relevance of this construction in CFT still waits to be uncovered.
Presumably, it corresponds to Knizhnik's idea [K]
of replacing the correlators of the analytic fields on
the covering of the Riemann surface, by the
correlators on the underlying Riemann surface with the insertions
of additional vertex operators.

Let us also remark, that quite analagous construction was
invented by Krichever in in [Kr] in connection
to the elliptic Calogero-Moser System.

\subsec{Degeneration of the spectral curve}

We shall adopt the
same definition of the spectral curve in the case of degenerate $\Sigma$.

Obviously, the normalization of
$S$ can be also decomposed
as the disjoint union of the
components $S_{\alpha}$,
labeled as the components  $\Sigma_{\alpha}$ and
$S_{\alpha}$ covers $\Sigma_{\alpha}$ with some fixed branching
at the points $x_{\alpha}^{i}$. Indeed, the behavior of $\phi_{\alpha}$
near the point $x_{\alpha}^{i}$ is known, since  the residue
is known. Let us fix the
conjugacy class of $p_{\alpha \beta}^{ij}$. Suppose, that
it has $k_{1}^{i}$ eigenvalues of multiplicity $1$, $k_{2}^{i}$ eigenvalkues
of multiplicity $2$, and so on. Since near the point $x_{\alpha}^{i}$
$\phi_{\alpha}$ behaves like:
$$
\phi_{\alpha} (x) \sim {p_{\alpha \beta}^{ij} \over{x - x_{\alpha}^{i}}}
$$
for appropriate $j$ and $\beta$, then $\lambda$ behaves like
$$
\lambda_{m}(x) \sim {p_{m} \over{x - x_{\alpha}^{i}}}
$$
where $p_{m}$ is the $m$'th eigenvalue of $p_{\alpha \beta}^{ij}$.
Following [BBKT], we can find
$$
N_{\alpha \beta}^{ij} = \sum_{m = 1}^{\infty} k_{m}^{i}
$$
points $P_{1}, \dots , P_{N_{\alpha \beta}^{ij}}$ above
$x_{\alpha}^{i}$, such that
the local parameteres
$Z_{1}, \dots, Z_{N_{\alpha \beta}^{ij}}$ near them are defined:
$$
Z_{m} = \lambda_{m}(x) - {p_{m} \over{x -  x_{\alpha}^{i}}}
$$

The discriminant $\Delta_{\alpha}$
of $\phi_{\alpha}$ is a meromorphic $N(N-1)$-differential on
${\Sigma}_{\alpha}$. At each point $x_{\alpha}^{i}$
it has a pole of the order
$$
o_{\alpha}^{i} = N^{2} - \sum_{m} k_{m}^{i} m^{2}.
$$
The zeroes of $\Delta_{\alpha}$
dertermine the branching points of the covering
$$
S_{\alpha} \to {\Sigma}_{\alpha}
$$
The number of the branching points equals, therefore, to
$$
2N(N-1)(g({\Sigma}_{\alpha})-1)
+ \sum_{i} o_{\alpha}^{i}
$$
The genus of $S_{\alpha}$ can be computed with the help of
Riemann-Hurwitz formula, which gives:
$$
g(S_{\alpha}) = 1 + N^{2}(g - 1) + {1\over{2}} \sum_{i} o_{\alpha}^{i}
$$

Now the Hamiltonian flow due to our Hamiltonians produces
a motion of the line bundle over $S_{\alpha}$ and it covers
the Jacobian of the completed curve $\bar S_{\alpha}$, therefore,
the coordinates of the particles
will be determined by the same
equation:
$$
{\theta}(\sum_{i} U^{i}t_{i} + Z_{0})
$$
as in the simplest one-punctured case.
Here $\theta$ is a $theta$-function on the
Jacobian of $\bar S_{\alpha}$, and ${\vec U}$ defines
an imbedding of the moduli space of the holomprphic
bundles over $\bar {\Sigma}_{\alpha}$ into the Jacobian, as we have described
it.

The details of the recontsruction of all angle-type variables will
be published elsewhere [NG3]. Remark, that this problem was
solved for one-punctured elliptic
curve for specific orbit in [BBKT].

\newsec{\bf  Formulas for general case - genus $g$ curve with $L$ punctures }

In this section we consider only one component $\Sigma$ of the stable curve.
We assume that it  has genus $g$ and $L$ punctures. We also
assume, that $\Sigma$ has no double points.

Using the formulas for the twisted meromorphic
forms on the curve, quoted in [I],
we can easily write down the formula for
the solution of the main equation
$$
[ d^{\prime \prime}_{A} , \phi ] + \sum_{i} p_{i} \delta^{2} (x_{i}) = 0
$$

In order to do this, we choose the following coordinatization
of the moduli space $\cal N$ of holomorphic bundles over $\Sigma$. Namely,
over the open dense subset of $\cal N$ one can parameterize
the holomorphic bundle by choosing a set of $g$ twists:
elements of the complex group $G$, assigned to the $A$ - cycles
of $\Sigma$. More  presicely, let us  fix the representatives $a_{k}$,
$k=1, \dots, g$,
of the $A$-cycles and let $\tilde \Sigma$ be the
surface $\Sigma$  with the small {\neib}s of $a_{k}$
removed. Topologically $\tilde \Sigma$ is a
sphere with $2g$ holes.

The boundary of the {\neib} of $a_{k}$
consists of two circles $a_{k}^{\pm}$.
In order to glue back the surface
$\Sigma$ one has to attach the projective transformations
${\gamma}_{k}$, which map $a_{k}^{+} \to a_{k}^{-}$.
These transformations generate Shottky group.
On the sphere one can
find such a gauge transformation $h$, that
$$
d^{\prime \prime}_{A} = h^{-1} \bar \partial h
$$
Obviously,
$$
h (g_{k}(x)) \vert_{a_{k}^{-}} = H_{k}(x) h(x) \vert_{a_{k}^{+}}
$$
where $H_{k}$ is a holomorphic $G$ -valued function,
defined in the vicinity of $a_{k}^{+}$. Generically one
can find constant representative of $H_{k}$ (this is
a Riemann-Hilbert problem).

Once such a gauge $h$ transformation is chosen, the
equation for $\phi$ can be restated in words as the following:
find a meromorphic form on $\tilde \Sigma$, which satisfy the following
requirements:

{\it in the vicinity of $x_{i}$: $\phi \sim {{p_{i}}\over{x - x_{i}}}$}

{\it twisting: ${\phi}({\gamma}_{k}(x)) d {\gamma}_{k}(x) =
Ad_{H_{k}} {\phi}(x) dx$}

The answer can be conviniently written in terms
of the Poincare seria ([I]):

Introduce
$$
\omega_{k}[x_{0}] \in \Omega^{1}(\C\P^{1}) \otimes End (Lie G),
$$
$$
\theta [x,x_{0}] \in \Omega^{1}(\C\P^{1}) \otimes End (Lie G),
$$
where $x, x_{0} \in \C\P^{1}$,
$$
\omega_{k}[x_{0}]_{z} dz =
\sum_{\gamma \in \Gamma} Ad(H_{\gamma}^{-1}) {\sl d} log
{{{\gamma}(z) - {\gamma}_{k}(x_{0})}\over{{\gamma}(z) - x_{0}}}
$$
$$
\theta [x, x_{0}]_{z} dz =
\sum_{\gamma \in \Gamma} Ad(H_{\gamma}^{-1}) {\sl d} log
{{{\gamma}(z) - x}\over{{\gamma}(z) - x_{0}}}
$$
where the sum runs over all elements of the Shottky group
(it is a free group with  $g$ generators - loops around $B$-cycles.
One assigns to the "word"
$$
\gamma = b_{i_{1}}^{n_{1}} b_{i_{2}}^{n_{2}} \dots  b_{i_{k}}^{n_{k}}
$$
an element of $G$:
$$
H_{\gamma} = H_{i_{1}}^{n_{1}} H_{i_{2}}^{n_{2}} \dots  H_{i_{k}}^{n_{k}} )
$$

In this formulas $x_{0}$ is an auxilliary point on the sphere.

Finally, the solution has the form:
$$
\phi [x_{0}] =
\sum_{k=1}^{g} \omega_{k}(w_{k}) [x_{0}] +
\sum_{i=1}^{L} \theta (p_{i}) [ x_{i}, x_{0} ]
$$

The momenta $w_{k}$ are defined from the condition:
$$
\int_{a_{k}} \phi = w_{k}
$$

Easy computation shows, that the symplectic form becomes
the sum of the symplectic forms on the orbits, attached to
the points $x_{i}$ and of the $g$ copies
of the Liouville form on the $T^{*}G$, where
the momentum for $H_{k}$ is  $w_{k}$.

The final remark concerns the vanishing of the residue
at the point $x_{0}$:
$$
\sum_{k} Ad(H_{k}^{-1})(w_{k}) - w_{k} + \sum_{i} p_{i} = 0
$$
This equation we met in the degenerate form in the section \S 4.1.
It has the meaning of the moment map for the action of $G$
on the product of $g$ copies of $T^{*}G$ and the coadjoint
orbits of $p_{i}$.

\newsec{ Applications to quantization}

It is straightforward to quantize our models. When the conjugacy classes
of $p_{\alpha \beta}^{ij}$ are fixed, their quantum counterparts
become simply the generators of the group, acting
in the correspondning representations of $G$. The  condition
on  the residues of $\phi_{\alpha}$ and $\phi_{\beta}$ at
the double point gets translated to  the fact, that
the representations, sitting at the points
$x_{\alpha}^{i}$ and $x_{\beta}^{j}$, belonging to one
double
point, are dual to each  other.

The pinched handle corresponds to the regular representation
of the group, and the corresponding generators
$p_{\alpha \alpha}^{ij}$ and $p_{\alpha \alpha}^{ji}$ are left- and right-
invariant vector fields on the group.

Then the  Schrodinger equations for the wavefunctions coincide
with the critical level Knizhnik-Zamolodchikov-Bernard equations [KZ],
[Be], [Lo], [EK1],[EFK].
The result of the quantization should follow from the degeneration
of the Beilinson-Drinfeld construction [Beil].

Also, it would be nice to realize the meaning of the
generalized KZ equations of [Ch] along the lines of our approach. As far
as it seems now, these equations are inspired by the occasional
fact, that the Hamiltonians we have written for
the punctured elliptic curve are almost symmetric
under the exchange :$z_{i} \leftrightarrow x_{a}$.

 Finally, note, that using the results of [I] one can easily
write down the quadratic Hamiltonians for an arbitrary curve (unfortunately,
at the moment only in terms of
the covering of the open dense subset of the actual phase space),
while [FV] allows one to get the expression
for the wave-functions of the elliptic Gaudin model in
terms of the solutions of the Bethe Ansatz-like equations.

When the paper was completed we have been notified about the recent
paper by B. Enriquez and V. Rubtsov [ER] on the related subject.
We would like to thank authors of [ER] for their comments.

\vfill\eject
\centerline{\bf REFERENCES}
\bigskip

\item{[BBKT]} {O. Babelon, E. Billey, I. Krichever, M.Talon, {\it Spin
generalization of the Calogero-Moser system and the Matrix KP equation},
hepth/9411160}
\item{[Beil]} {A. Beilinson, V. Drinfeld, {\it Quantization
of Hitchin's fibration and Langlands program},
{\rm A. Beilinson's lectures
at IAS, fall 1994}}
\item{[Be]} {D. Bernard, Nucl. Phys. {\bf B}303 (1988)77, {\bf B}309 (1988),
14}
\item{[C]} {F. Calogero, J.Math.Phys. {\bf 12} (1971) 419;}
\item{[Ch]} {I.Cherednik,  {\it Difference-elliptic operators and root
systems, hep-th/9410188}}
\item{[EF]} {P.Etingof, I.Frenkel, {\it Central extensions of current groups
in two dimensions}, hepth/9303047}
\item{[EK1]}
{P.Etingof, {\it Quantum integrable systems and representations of
Lie algebras}, hepth/9311132}
\item{[EK2]} {P.Etingof, A.Kirillov, Jr., {\it On the affine
analogue of Jack's and Macdonald's polynomials}, Yale preprint, 1994}
\item{[EFK]} {P. Etingof, I.Frenkel, A. Kirillov-Jr., {\it Spherical functions
on affine Lie groups}, Yale preprint, 1994}
\item{[FV]} {G. Felder, A. Varchenko, {\it Integral formula for the
solutions of the elliptic KZB  equations}, hepth/9503001}
\item{[NG1]} {A. Gorsky, N. Nekrasov, {\it Elliptic Calogero-Moser System from
Two Dimensional Current Algebra}, hepth/9401021}
\item{[NG2]} {A.Gorsky, N.Nekrasov, {\it Relativistic Calogero-Moser model as
gauged WZW theory}, Nucl.Phys. {\bf B} 436 (1995) 582-608}
\item{[NG3]} {N. Nekrasov, A. Gorsky , {\it Integrable systems on moduli
spaces}, in preparation}
\item{[FF]} {B. Feigin , E. Frenkel, N. Reshetikhin, {\it Gaudin Model,
Critical Level and Bethe Ansatz, CMP {\bf  166} (1995), 27-62}
\item{[G]} {M. Gaudin, Jour. Physique, {\bf 37} (1976), 1087-1098}
\item{[H]} {N. Hitchin, Duke Math. Journal, Vol. 54, No. 1 (1987)}
\item{[I]} {D. Ivanov,
{\it Knizhnik-Zamolodchikov-Bernard equations on Riemann surfaces.},
hep-th/9410091}
\item{[KKS]} {D.Kazhdan, B.Kostant and S.Sternberg, Comm. on Pure and
Appl. Math., vol. {\bf XXXI}, (1978), 481-507}
\item{[KZ]}{V. Knizhnik, A. Zamolodchikov, Nucl. Phys. {\bf B}247 (1984) 83}
\item{[K]} {V. Knizhnik, {\it Analytic fields on Riemann surfaces}, CMP
{\bf 112}, (1987), 567}
\item{[Lo]}{A. Losev, {\it Coset construction and Bernard Equations},
CERN-TH.6215/91}
\item{[OP]} {M.Olshanetsky, A.Perelomov, Phys. Rep. {\bf 71} (1981), 313}
\item{[M]} {J. Moser, Adv.Math. {\bf 16} (1975), 197-220; }
\item{[Kr]} { I. Krichever, Funk. Anal. and Appl., {\bf 12}
(1978),  1, 76-78; {\bf 14} (1980), 282-290}
\item{[RS]} {S.N.M. Ruijsenaars,
H. Schneider, Ann. of Physics {\bf 170} (1986),
 370-405}
\item{[R]} {S. Ruijsenaars, CMP {\bf 110}  (1987), 191-213 }
\item{[S]} {B. Sutherland, Phys. Rev. {\bf A5} (1972), 1372-1376;}
\item{[V]} {E. Verlinde, Nucl. Phys. {\bf B} 300 (1988) 360}
\item{[ER]} {B. Enriquez, V. Rubtsov, {\it Hitchin systems, higher
Gaudin operators and r-matrices}, alg-geom/9503010}
\vfill\eject
\bye